\documentclass[prb,twocolumn]{revtex4}
\usepackage{epsfig}
\usepackage{dcolumn}

\usepackage[latin1]{inputenc}
\usepackage[T1]{fontenc}

\begin{document}
\newcommand{\degree}{$^{\rm\circ} $}
\newcommand{\pcite}{\protect\cite}
\newcommand{\pref}{\protect\ref}
\newcommand{\rfg}[1]{Fig. \ref{F#1}}
\newcommand{\rtb}[1]{Table \ref{T#1}}

\title{Single Stranded Breaks Relax Intrinsic Curvature in DNA}

\author{Dimitri E. \surname{Kamashev}}
\affiliation{CNRS UPR9051, H\^opital St. Lois, 1
av.  C. Vellefaux, 75475 Paris Cedex 10, France.}

\author{Alexey K. \surname{Mazur}}
\email{alexey@ibpc.fr}
\altaffiliation{FAX:+33[0]1.58.41.50.26}
\affiliation{Laboratoire de Biochimie Th\'eorique, CNRS UPR9080,
Institut de Biologie Physico-Chimique,
13, rue Pierre et Marie Curie, Paris,75005, France}


\begin{abstract}

The macroscopic curvature of double helical DNA induced by regularly
repeated adenine tracts (A-tracts) is well-known but still puzzling.
Its physical origin remains controversial even though it is perhaps
the best-documented sequence modulation of DNA structure. The recently
proposed compressed backbone theory (CBT) suggested that the bending
can result from a geometric mismatch between the specific backbone
length and optimal base stacking orientations in B-DNA. It predicted
that the curvature in A-tract repeats can be relaxed by introducing
single stranded breaks (nicks). This effect have not been tested
earlier and it would not be accounted for by alternative models of DNA
bending.  This paper checks the above prediction in a combined
theoretical and experimental investigation. A series of nicked DNA
fragments was constructed from two mother sequences with identical
base pair composition, one including A-tract repeats and the other
being random.  The curvature was tested experimentally by gel mobility
assays and, simultaneously, by free molecular dynamics (MD)
simulations.  Single stranded breaks produce virtually no effect upon
the gel mobility of the random sequence DNA. In contrast, for nicked
A-tract fragments, a regular modulation of curvature is observed
depending upon the position of the strand break with respect to the
overall bend. As predicted by CBT, the curvature is reduced, with
maximal relaxation observed when the nick occurs inside A-tracts.
These experimental results are partially reproduced in MD simulations
of nicked A-tract repeats. Analysis of computed curved DNA
conformations reveals modulations of local backbone length as measured
by distances between some sugar atoms, with maximal backbone
compression in the narrowing of the minor groove inside A-tracts. The
results lend additional support to CBT versus alternative mechanisms
of intrinsic curvature in DNA.\end{abstract}


\maketitle

\section{Introduction}

The intrinsic curvature in DNA was discovered about twenty years ago
for regular repeats of $\rm A_nT_m,\ with\ n+m>3$, called A-tracts
\cite{Marini:82,Wu:84,Hagerman:84}. Several profound reviews exist of
the large volume of related data accumulated during the last decades
\cite{Diekmann:87a,Hagerman:90,Crothers:90,Crothers:92,Olson:96,Crothers:99}.
Due to its outstanding role in genome functioning and a very
controversial experimental behavior, the sequence dependent DNA
bending continues to attract great attention. The hottest point of
debates is the molecular mechanism of this phenomenon.  Already in the
seventies, it was realized that the DNA double helix is not a rigid
cylinder, but is bendable depending upon direction and the base pair
sequence \cite{Namoradze:77,Zhurkin:79}. By postulating that stacking
in ApA steps is intrinsically non-parallel the wedge model predicted
that A-tracts repeated in phase with the helical screw should bend
DNA\cite{Trifonov:80}. This proved true, but the model later appeared
unsatisfactory. It turned out that X-ray DNA structures are often bent
in random sequences whereas A-tracts are exceptionally straight
\cite{Calladine:88,Maroun:88,Dickerson:94}. With the growing number of
sequences checked, non-zero wedges had to be introduced for all base
pair steps \cite{DeSantis:90,Bolshoy:91,Liu:01}, nevertheless,
experimental counter-examples were eventually found where bending
could not result from accumulation of wedges
\cite{Dlakic:96a,Dlakic:98}. Several theories tried to reveal other
physical factors involved in DNA bending along with base stacking. The
most popular junction model proposed by the Crother's
group\cite{Levene:83,Wu:84} originated from an idea that a bend should
occur when two different DNA forms are stacked \cite{Selsing:79}. If
poly(dA).poly(dT) double helix had a special B' form as suggested by some data
\cite{Alexeev:87} the helical axis should be kinked when an A-tract is
interrupted by a random sequence.  Drew and Travers
\cite{Drew:84,Drew:85} apparently were the first to notice that
narrowing of both DNA grooves at the inner edge of a bend is a
necessary and sufficient condition of bending.  They, and later
Burkhoff and Tullius \cite{Burkhoff:87}, considered the preference of
narrow and wide minor groove profiles by certain sequences as the
possible original cause of this effect. Similar ideas were considered
within the context of the junction model\cite{Sprous:99}. In addition,
the possible involvement of solvent counterions in DNA bending was
discussed long ago\cite{Levene:86}, and it has been extensively
studied in the recent years\cite{Young:97a,Rouzina:98,Hud:01}.
Detailed analyses of all these theories including comparison with
experimental data can be found in the recent literature
\cite{Olson:96,Crothers:99,Mzpre:02}. Here we note only that, in spite
of all efforts, the controversies remain and continue to accumulate
\cite{Merling:03,Hud:03}.

Conformational calculations were always intensively used for testing
various hypothesis concerning DNA bending
\cite{Levene:83,Koo:90,Zhurkin:79,Ulyanov:84,Sanghani:96,Kitzing:87,Chuprina:88,Zhurkin:91}.
The progress in methodology of molecular dynamics (MD) simulations of
nucleic acids \cite{Cheatham:00} recently made possible realistic
modeling of A-tract DNA fragments
\cite{Young:98,Sprous:99,Sherer:99,Strahs:00}. In straightforward
simulations of bending, the initial straight symmetrical DNA is
allowed to bend spontaneously with no extra forces applied, that is
due to generic atom-atom interactions
\cite{Young:98,Sprous:99,Mzjacs:00,Mzjbsd:01,Mzpre:02}. Such
computational experiments are very demanding because they involve
relatively long DNA fragments and require multi-nanosecond (at least)
trajectories to reveal reliable trends toward bent states.
Nevertheless, for preselected A-tract sequences, it was possible to
demonstrate spontaneous emergence of intrinsic curvature in the course
of simulations as well as attracting character of bent states, which
is necessary for static curvature. These simulations resulted in a new
hypothesis of the physical origin of intrinsic curvature which we
refer to as the compressed backbone theory
(CBT)\cite{Mzjacs:00,Mzjbsd:01,Mzpre:02}.

In a certain sense, the CBT continues the line of earlier views that
attributed the origin of bending to the properties of the DNA
backbone\cite{Drew:84,Drew:85,Burkhoff:87,Sprous:99}.  It postulates
that the intrinsic curvature in B-DNA results from a geometric
mismatch between the length of the sugar-phosphate backbone and the
base pair stacking.  Starting from the first published drawing of the
double helix \cite{Watson:53}, its two backbone strands are viewed as
regular spiral traces that wind along the surface of the cylindrical
B-DNA core formed by stacked base pairs. Because regular spirals
represent the shortest lines that join two points on a cylindrical
surface, and because the backbone strands are covalently linked to
bases, an ideal B-DNA can exist only if the average backbone length
either matches exactly or is shorter than that dictated by the base
pair stacking \cite{Mzjacs:00}. The CBT assumes the opposite.  It
postulates that, under physiological conditions, the equilibrium
specific length of the B-DNA backbone, considered as a restrained
polymer attached to a cylindrical surface, is slightly longer that in
the canonical B-form. In physics, similar mismatching relationship is
often called geometric frustration \cite{Ediger:00}.  The backbone
tries to expand and "pushes" stacked bases while the stacking
interactions oppose this. Eventually, the backbone finds a
"compromise" by deviating from its regular spiral trace, which causes
quasi-sinusoidal modulations of the DNA grooves.  Concomitant base
stacking perturbations cause local bends that accumulate to
macroscopic static curvature when the period of these modulations
corresponds to an integral number of helical turns. The CBT considers
B-DNA double helices with any sequences and it may have a number of
interesting consequences.  It suggests, for example, that, even under
normal temperature, the B-DNA double helix can occur in a glass-like
state characterized by very long relaxation times \cite{Mzpre:02}. The
latter prediction agrees with some recent experimental data
\cite{Song:90,Okonogi:99,Brauns:02}. Under certain assumptions, one
can even imagine the possibility of a glass-like aging of a single
DNA molecule {\em in vivo}, with direct relation to the problems of
biological aging.

One of the possible experimental tests of CBT is based upon the
predicted effect of single-stranded breaks (nicks) upon the intrinsic
curvature \cite{Mzpre:02}. Reversible single chain breaks in DNA occur
during diverse biochemical processes, including general and
site-specific recombination, replication, and DNA repair. Therefore,
structural perturbations due to nicks and single-stranded "gaps" have
been earlier characterized in a number experimental studies
\cite{Thomas:56,Hays:70,Shore:83,Hsieh:89,Pieters:89,Snowden:90,Aymami:90,Rentzeperis:93,LeCam:94,Mills:94,Hagerman:96,Furrer:97,Lane:97,Singh:97,Roll:98,Kozerski:01,Kuhn:02}.
According to all these data, nicks perturb the free DNA very little.
The corresponding X-ray structures\cite{Aymami:90} are straight and
similar to those of intact duplexes with analogous sequences. This
agrees with early physicochemical tests
\cite{Thomas:56,Hays:70,Shore:83} as well as NMR studies in solution
\cite{Pieters:89,Snowden:90,Singh:97,Roll:98,Kozerski:01}. The main
detectable effect of a nick is local melting or fraying, especially at
low ionic strength, mild denaturating conditions, or elevated
temperatures. This increases the isotropic flexibility of DNA and
reduces its gel mobility \cite{LeCam:94,Mills:94,Furrer:97,Kuhn:02}.
In normal conditions the nick fraying is small and it almost
undetectable below 10\degree C\cite{Mills:94,Kuhn:02}.

Based upon these experimental data and the earlier views of the origin
of intrinsic curvature one should expect that nicks in phased A-tract
sequences can either increase bending or have zero effect. The
increase may be expected if the curvature is caused by external
forces, as in electrostatic counterion models, because the flexible
hinge at a nick can allow these forces to increase the bend angle in
their direction. A zero effect is expected for the wedge and junction
models because nicks arguably do not affect the A-tract structures
\cite{Aymami:90} and do not cause wedges detectable in gel migration
assays \cite{Mills:94}. In contrast, if the intrinsic curvature is
really caused by the backbone compression, single-stranded breaks
should relax it, which should increase the gel mobility of curved DNA
fragments. This effect is opposite to that of the increased isotropic
flexibility, moreover, the latter can be eliminated by reducing the
temperature so that these two phenomena cannot be confused. In
addition, because the degree of the compression should vary regularly
between the inner and outer edges of the bent DNA, relaxation produced
by a single nick is expected to change systematically according to its
position with respect to the overall bend.

Here we present the results of experimental and simulation studies
carried out according to the above plan. Two 35-mer sequences from our
recent comparative investigation\cite{Mzpre:02} were used for
constructing two series of nicked DNA fragments. The first sequence
contained an A-tract repeat motif that was preselected in MD
simulations to reproducibly induce strong static curvature. The second
sequence was ``random'', but with the same base pair composition.
Backbone breaks were introduced in one strand at different portions to
span approximately one helical turn. The curvature was probed by PAGE
as well as long-time MD simulations. In experiment, a noticeable
increase in the gel mobility of the nicked A-tract repeat is observed
depending upon the nick position. In contrast, no such effect is found
for the random fragment. MD simulations fail to reproduce experimental
effects qualitatively, but qualitatively they also confirm that single
stranded breaks interfere with the A-tract induced curvature. The
results qualitatively agree with CBT, but they cannot be accounted for
by other models.


\section{Materials and Methods}

\subsection{Oligonucleotides and construction of 5'-labeled DNA
probes}

The double stranded DNA containing four A-tracts flanked
poly(dA).poly(dT) termini were constructed by annealing the synthetic
oligonucleotide B (the top line in \rtb{alseq}) with its complementary
(bottom) strand. Double stranded DNA fragments containing single
stranded breaks (nicks) were constructed by annealing two shorter
oligonucleotides (\rtb{alseq}) with the same bottom strand. In the
same way a series of nicked duplexes was constructed from the
reference random sequence fragment S, with its base pair composition
identical to that of B. The sequences of the oligonucleotides used for
the construction of the DNA fragments are assembled in \rtb{alseq}.
The fragment codes in this table use the following mnemonics. The
capital 'B' and 'S' stand for "bent" and "straight". The small letter
prefix 'n' stands for "nick". The numbers indicate the shift of the
nick position with respect to the center of the sequence.

In all the constructs, one of the oligonucleotides annealed was
labeled with T4 polynucleotide kinase and [$^{32}$-P]-ATP. In most
cases the label was attached to the 5'-end of the bottom (continuous)
oligonucleotide whereas the two strands at the nick position were
terminated with 3'OH and 5'OH groups. For some sequences,
5'-phosphorylated nicks were additionally constructed to check the
effect of phosphate charges. In this case the continuous strand
carried 5'-OH while [$^{32}$-P] was introduced in the corresponding
shorter partner.  The annealing was carried out by incubating 300 nM
of the unlabeled oligonucleotide(s) with 30 nM of end-labeled ones for
3 min at 80\degree C in 20 mM Tris-HCl (pH 8.0), 150 mM NaCl, and then
allowing them to cool slowly. To distinguish between the two nick
series, in the text, we add small letter suffixes 'o' and 'p' to the
codes given in \rtb{alseq}. Thus the whole two nick series of the bent
DNA are referred to as nBo and nBp, respectively, while the individual
nicked fragments are denoted as nB-2o, nB-2p, etc.

\begin{table*}[t]\caption{\label{Talseq}Construction of DNA fragments.
A-tracts are boldfaced and nick positions are marked by double quotes.}
\begin{ruledtabular} \begin{tabular}[t]{|cc|}
Code & Sequence\\
\hline
B    & A$_{18}$-{\bf AAAAT}AGGCT{\bf ATTTT}AGGCT{\bf ATTTT}AGGCT{\bf ATTTT}-T$_{18}$\\
nB-9 & A$_{18}$-{\bf AAAAT}AGG"CT{\bf ATTTT}AGGCT{\bf ATTTT}AGGCT{\bf ATTTT}-T$_{18}$\\
nB-6 & A$_{18}$-{\bf AAAAT}AGGCT{\bf A"TTTT}AGGCT{\bf ATTTT}AGGCT{\bf ATTTT}-T$_{18}$\\
nB-4 & A$_{18}$-{\bf AAAAT}AGGCT{\bf ATT"TT}AGGCT{\bf ATTTT}AGGCT{\bf ATTTT}-T$_{18}$\\
nB-2 & A$_{18}$-{\bf AAAAT}AGGCT{\bf ATTTT}"AGGCT{\bf ATTTT}AGGCT{\bf ATTTT}-T$_{18}$\\
nB+0 & A$_{18}$-{\bf AAAAT}AGGCT{\bf ATTTT}AG"GCT{\bf ATTTT}AGGCT{\bf ATTTT}-T$_{18}$\\
nB+2 & A$_{18}$-{\bf AAAAT}AGGCT{\bf ATTTT}AGGC"T{\bf ATTTT}AGGCT{\bf ATTTT}-T$_{18}$\\
nB+4 & A$_{18}$-{\bf AAAAT}AGGCT{\bf ATTTT}AGGCT{\bf A"TTTT}AGGCT{\bf ATTTT}-T$_{18}$\\
nB+6 & A$_{18}$-{\bf AAAAT}AGGCT{\bf ATTTT}AGGCT{\bf ATT"TT}AGGCT{\bf ATTTT}-T$_{18}$\\
nB+11& A$_{18}$-{\bf AAAAT}AGGCT{\bf ATTTT}AGGCT{\bf ATTTT}AGG"CT{\bf ATTTT}-T$_{18}$\\
S    & A$_{18}$-TTAGATAGTATGACTATCTATGATCATGTATGATA-T$_{18}$\\
nS-9 & A$_{18}$-TTAGATAG"TATGACTATCTATGATCATGTATGATA-T$_{18}$\\
nS-6 & A$_{18}$-TTAGATAGTAT"GACTATCTATGATCATGTATGATA-T$_{18}$\\
nS-4 & A$_{18}$-TTAGATAGTATGA"CTATCTATGATCATGTATGATA-T$_{18}$\\
nS-2 & A$_{18}$-TTAGATAGTATGACT"ATCTATGATCATGTATGATA-T$_{18}$\\
nS+0 & A$_{18}$-TTAGATAGTATGACTAT"CTATGATCATGTATGATA-T$_{18}$\\
nS+2 & A$_{18}$-TTAGATAGTATGACTATCT"ATGATCATGTATGATA-T$_{18}$\\
nS+4 & A$_{18}$-TTAGATAGTATGACTATCTAT"GATCATGTATGATA-T$_{18}$\\
nS+6 & A$_{18}$-TTAGATAGTATGACTATCTATGA"TCATGTATGATA-T$_{18}$\\
nS+11& A$_{18}$-TTAGATAGTATGACTATCTATGATCATG"TATGATA-T$_{18}$\\
\end{tabular}
\end{ruledtabular} \end{table*}

\subsection{Gel mobility assays}

The mobility of the DNA fragments was analyzed in 12\% gels (acrylamide to
bis-acrylamide, 29:1) buffered with 90 mM Tris-borate, 1 mM EDTA, pH
8.6. In order to check the effect of Mg$^{2+}$ ions the buffer was
supplemented with 10 mM MgCl$_2$ in the absence of EDTA.  Gels were pre-run
under constant power until stabilization of the current. Labeled DNA in a
buffer containing 20 mM Tris-HCl (pH 8.0), 50 mM NaCl, 4\% Phicoll-400
and xylencianol was loaded onto the gel. The electrophoresis was
performed under constant voltage and constant temperature of 4\degree
C. The dried gels were exposed to storage phosphor screens and
visualized on a 400S PhosphorImager (Molecular Dynamics).

\subsection{Calculations}

MD simulations were carried out for the central 35 base pairs of
six nicked DNA fragments in \rtb{alseq} constructed from the A-tract
repeat sequence (nB-4o, ..., nB+6o). The corresponding calculations
for the two mother fragments without nicks (B and S) were reported
earlier\cite{Mzpre:02}. For consistency with the previous
studies\cite{Mzjacs:00,Mzjbsd:01,Mzpre:02}, we used the AMBER98
\cite{Cornell:95,Cheatham:99} force field and TIP3P water
\cite{TIP3P:}. Trajectories generally started from the fiber canonical
B-DNA conformation\cite{Arnott:72} and were continued to 20 ns. In a
few cases, dynamics was also run starting from the canonical A-DNA
conformation. In total, these simulations sampled from more than 0.2
ms of all-atom dynamics for nicked and intact 35-mer DNA fragments
with the same A-tract repeat sequence.

Molecular dynamics simulations were carried out with the internal
coordinate molecular dynamics (ICMD) method \cite{Mzjcc:97,Mzbook:01}
adapted for DNA \cite{Mzjacs:98,Mzjchp:99} with the time step of 0.01
ps. In this approach, the DNA molecule has all bond length and almost
all bond angles fixed at their standard values. The only variable bond
angles are those centered at the sugar C1',...,C4', and O4' atoms,
which assures the flexibility of the furanose rings. In contrast,
bases, thymine methyls, and phosphate groups move as articulated rigid
bodies, with only rotations around single bonds allowed. The highest
frequencies in thus obtained models are additionally balanced by
increasing rotational inertia of the lightest rigid bodies as
described earlier\cite{Mzjpc:98,Mzjacs:98}. The possible physical
effects of the above modifications have been discussed elsewhere
\cite{Mzbook:01,Mzctpc:01}.

The so-called "minimal model" of B-DNA was used
\cite{Mzjacs:98,Mzjcc:01}. It includes only a partial hydration shell
and treats counterion and long range solvation effects implicitly by
reducing phosphate charges to -0.5 and applying linear scaling of
Coulomb forces. This model produces B-DNA structures very close to
experimental data and has no other bias towards bent or non-bent
conformations except the base pair sequence.  The interest to implicit
solvation in DNA simulations is long standing and it continues in the
literature \cite{Wang:02b}. Advantages as well as limitations of such
approaches have been recently reviewed \cite{Cheatham:00}.

The minimal model is not meant to be generally applicable and we have
chosen it for the present studies because of the following
considerations. We assume that effects it ignores, like sequence
specific counterion binding, play some role, but are not critical for
the A-tract induced curvature. We have shown earlier that, in these
conditions, the A-tract repeat motif we are using has a bent state
with very strong attracting properties that allow one to observe
spontaneous transitions to stable bent conformations in the course of
MD \cite{Mzjacs:00,Mzpre:02}. This property is rather exceptional and
only due to it one could hope to obtain useful information from a
comparison between nicked and intact DNA in MD. We also carried out
several simulations for the mother 35-mer and 25-mer A-tract fragments
by the Particle-Mesh Ewald method\cite{Essmann:95} with full solvation
and periodical boundaries. The corresponding trajectories were
continued to 5-10 ns, but they showed much less demonstrative bending
dynamics. The origin of this difficulty is currently unclear, but it
agrees with recent reports by some other authors \cite{Cheatham:01}.
Finally, the minimal model provides for substantial savings in
computations. The saving factor is critical because DNA bending is
detectable only in relatively long fragments and it is likely to
involve very slow motions. In spite of a few reported analogous
simulations using full-scale solvation \cite{Young:98,Sprous:99} such
calculations would be too costly for the volume of sampling reached
here.

\section{Results and Discussion}

\subsection{Construction of DNA Fragments}

The A-tract motif $\rm AAAATAG$ originally attracted our attention in
MD simulations of the natural DNA fragment taken from the first curved
DNA locus studied {\em in vitro}\cite{Wu:84,Mzjbsd:01}. The central
35-mer A-tract repeat for the bent fragments in \rtb{alseq} was
constructed by repeating this motif four times and it had to be
inverted to make the two DNA termini symmetrical. Such inversion
should not affect bending, \cite{Koo:86} but is essential for
simulations because the 3'- and 5'-end A-tracts may represent
qualitatively different boundaries. In repeated simulations with this
and similar A-tract fragments, the static curvature emerged
spontaneously and it became more evident as the chain length increased
\cite{Mzjacs:00,Mzpre:02}. To obtain a reference non-A-tract DNA, we
have re-shuffled manually base pairs of the A-tract repeat. We
preferred this randomized sequence to commonly used GC-rich straight
fragments in order to keep the base pair content identical and reduce
the noise that could cause small variations in gel mobility. In order
to amplify the PAGE resolution of similarly bent nicked DNA, the
35-mer fragments were extended to 71 bp by adding poly(dA).poly(dT)
tails. The tails continue the mother 35-mer A-tract repeat sequence
smoothly, which should reduce the possible perturbations that could
affect the comparison of PAGE results with MD simulations that
involved only the central 35-mer nicked fragments.

\subsection{Relative PAGE Mobilities}

A representative PAGE plate of the random sequence series of
non-phosphorylated nicks is shown in \rfg{nSo}. All nicked DNA exhibit
similar mobilities close to that of the reference straight fragment. A
weak retardation effect distinguishable for some nick positions is at
the limit of experimental accuracy. Qualitatively similar patterns,
but with stronger retardation were earlier reported for nicked DNA
fragments under elevated temperature and/or mild denaturating
conditions \cite{LeCam:94,Mills:94,Kuhn:02}. We suppose, therefore,
that some reduction of mobility in \rfg{nSo} results from the
isotropic flexibility at nick positions that is strongly reduced at
4\degree C. This small retardation should be taken into account in the
interpretation of other tests below.

\begin{figure}
\centerline{\psfig{figure=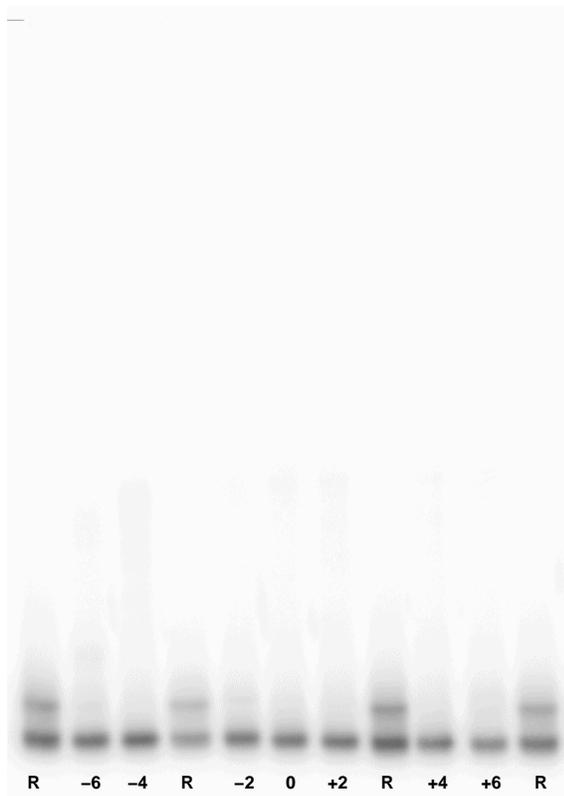,width=8.2cm,angle=0.}}
\caption{\label{FnSo}
PAGE analysis of comparative mobilities of nicked DNA constructed from
the random fragment. The contents of the lanes is as follows (see
\rtb{alseq} and Methods for fragment codes). The lanes marked 'R'
("reference") contain a mixture of intact duplexes S and B.  The
series of lanes '-6', ..., '+6' contain nS-6o, ..., nS+6o,
respectively. }\end{figure}

\rfg{nBo} shows the results obtained in the same conditions for the
A-tract nicks. The pattern exhibited here is evidently more complex
than that in \rfg{nSo}. On the one hand, for the central nick
positions, the mobilities are close to that of the mother fragment,
which is qualitatively similar to \rfg{nSo}.  However, when the nick
is moved farther from the center, the PAGE mobility grows and becomes
higher than that of the mother curved fragment. It reaches maximum
values for nB-4o and nB+6o, but for nB-6o it is reduced again.  As a
result, the mobilities of nB-6o and nB-4o appear close to those of
nB+4o and nB+4o, respectively, and the overall pattern of nick bands
in \rfg{nBo} appears skewed-sinusoidal, with the period corresponding
to that of the double helix. The increase of PAGE mobility indicates
reduced DNA curvature and this effect can well be due to the
relaxation predicted by CBT.  The amplitude of the reduction is very
significant since it exceeds 30\% of the curvature difference between
the S and B fragments produced by four A-tracts. This mobility profile
suggests that the backbone compression is maximal within A-tracts and
minimal between them. Although this specific phasing is right opposite
to our first guess\cite{Mzpre:02} it is accounted for by CBT and not
by other theories as we discuss further below. However, there are a
few alternative interpretations that should be considered first.

\begin{figure}
\centerline{\psfig{figure=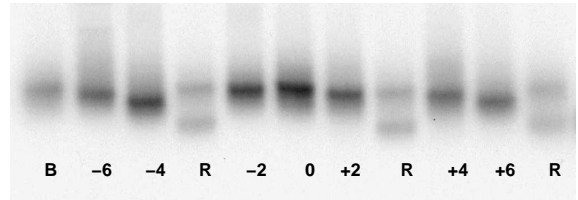,width=8.2cm,angle=0.}}
\caption{\label{FnBo}
PAGE analysis of comparative mobilities of nicked DNA constructed from
the A-tract repeat fragment. The contents of the lanes is as follows
(see \rtb{alseq} and Methods for fragment codes). The lanes marked 'B'
and 'R' ("reference") contain the intact B and its mixture with S,
respectively.  The series of lanes '-6', ..., '+6' contain nB-6o, ...,
nB+6o, respectively.}\end{figure}

The first evident factor that should be checked is the perturbed
balance of phosphate charges. The two nick series displayed in
\rfg{nSo} and \rfg{nBo} have one phosphate group less than the
reference fragments B and S. Consequently, their mobilities may be
uniformly reduced by around 0.7\%, which would mean that the apparent
reduction of curvature in the nicked A-tract fragments nB-4o and nB+6o
is even stronger than it is seems from \rfg{nBo}. This effect,
however, is not distinguishable in \rfg{nSo} suggesting that it is too
small and may be safely neglected. The influence of phosphate charges
can be more subtle, however. The electrostatic models of DNA bending
assume that it is the broken balance of phosphate repulsion at the
opposite DNA sides that forces it to bend \cite{Hud:01}. If
the A-tract curvature had an electrostatic origin, the phosphate
"hole" at the nick could perturb it by adding a local bend in a
direction that should rotate as the "hole" is moved along the DNA
chain. Like that the integral curvature would be deviated and
partially compensated in a way compatible with the results shown in
\rfg{nBo}.

To check this possibility, similar experiments were carried out
for a series of analogous phosphorylated nicks. Their relative PAGE
mobilities are compared with other results in \rfg{almob}. The
mobility coefficients Q$_m$ used in this plate were computed as
follows. The Q$_m$ value of the S fragment (see \rtb{alseq}) is always
the largest in the gel and it is arbitrarily assumed to equal 100. The
Q$_m$ of fragment B is assumed to equal 0. For any given band its
Q$_m$ is promotional to the distance from the fastest S band in the
same gel and it is estimated relative to the distance between B and S.
If a nicked fragment migrates slower than the B-band in the same gel
its Q$_m$ is negative. In contrast, if it is faster, the corresponding
Q$_m$ is positive below 100. \rfg{almob} displays thus obtained Q$_m$
values for all DNA fragments used in the present study.

\begin{figure}
\centerline{\psfig{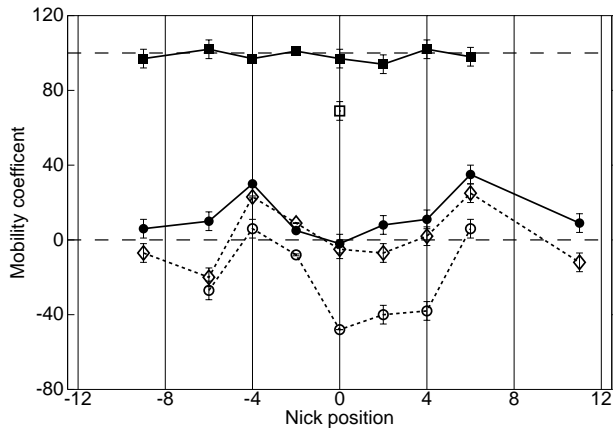}}
\caption{\label{Falmob}
Quantitative comparison of relative mobilities of DNA fragments listed
in \rtb{alseq}. The method of calculation of the mobility coefficients
Q$_m$ is explained in the text. The reference mobilities of fragments
B and S are indicated by horizontal dashed lines.  The error bars
indicate the estimate of experimental errors obtained from variation of
Q$_m$ in repeated runs.} \end{figure}

With free 5'-phosphates added to the nick sites, the phosphate "holes"
are quenched or even inverted, which should drastically change the
observed curvature modulations.  The results in \rfg{almob}, however,
show the opposite. The original skewed-sinusoidal dependence is
persistent.  At the same time, the phosphorylated nicks exhibit
uniformly reduced mobilities, which agrees with earlier
reports\cite{LeCam:94,Mills:94} and should be attributed to the
increased strand fraying in nicks.  The additional phosphate charge
increases the inter-strand electrostatic repulsion at single
stranded breaks, which should enhance the fraying effect as discussed
in detail elsewhere \cite{Furrer:97}. Whatever the reason of the
uniformly reduced mobility, however, \rfg{almob} clearly shows that
the above hypothesis of electrostatic curvature compensation can be
ruled out.

This figure also shows the profile of mobilities of the nBo series in
the presence of Mg$^{2+}$ ions. The Mg$^{2+}$ ions are long known to
affect the DNA curvature by increasing or reducing it depending upon
the sequence \cite{Diekmann:85,Diekmann:87c}. In our case the
curvature was increased as judged from the absolute PAGE separation of
fragments B and S.  It is seen in \rfg{almob} that the profile of the
nick mobilities in the presence of Mg$^{2+}$ ions remains
qualitatively similar, but some nicks migrated slower than the mother
B fragment.  Interpretation of this effect depends upon the mechanism
by which Mg$^{2+}$ increase the A-tract curvature. If the additional
bend is caused by the electrostatic attraction due to Mg$^{2+}$
positioning between the phosphate groups in the narrowings of the
major groove than the gain in the curvature may be larger when the DNA
is nicked in the widening of the minor groove at the opposite side of
the bend. This interpretation would agree with \rfg{almob} because the
negative Q$_m$ are observed for nick positions outside A-tracts in the
probable widenings of the minor groove. It is also possible, however,
that Mg$^{2+}$ ions somehow increase the nick fraying especially when
the nick position approaches the poly(dA).poly(dT) tails. In this
case, the negative Q$_m$ values would have the same origin as for the
phosphorylated nicks.


The results in \rfg{nBo} can also be interpreted with the wedge model
of DNA bending by assuming that nicks have a special structure with a
wedge in a fixed local direction.  The overall curvature is reduced
when this wedge direction is opposite to that of the initial bend and
increased in the opposite orientation.  Similarly to the "phosphate
hole" model above, skewed-sinusoidal modulations of curvature should
occur as the nick site is translated along DNA. Note that this wedge
should have a very large angle since it compensates in nB-4o and nB+6o
more than 30\% of the curvature due to four phased A-tracts. A strong
wedge like that should result in an increased curvature in nB+0o as
well as in the nSo series in \rfg{nSo}, which is not seen. This
interpretation also would not agree with earlier studies of nick
structures and gel mobilities \cite{Aymami:90,Mills:94,Furrer:97}.
However, earlier PAGE studies were carried out at higher temperatures
where the isotropic flexibility of nicks could, in principle, dominate
all other effects.  Also, the non-linear dependence of the gel
mobility upon the curvature may be such that the acceleration effect
of the nick wedge in \rfg{nBo} is more pronounced than the expected
retardation for nB+0o and the nSo series. Therefore, in order to check
the above explanations, double nicked A-tract fragments were
constructed, with the two sites separated by an odd number of helical
half-turns so that the hypothetical wedges had opposite directions and
should have compensated one another. The PAGE mobilities of double
nicked fragments are compared in \rfg{2n}.  These data evidently
disagree with the foregoing wedge interpretation of mobility
modulations. For both nB-4o and nB+6o, the addition of a second nick
at 1.5 helical turn neither cancel nor even reduce the corresponding
band shifts with respect to \rfg{nBo}. As expected, analogous
constructs for fragment S all show PAGE mobilities similar to that of
the intact duplex.

\begin{figure}
\centerline{\psfig{figure=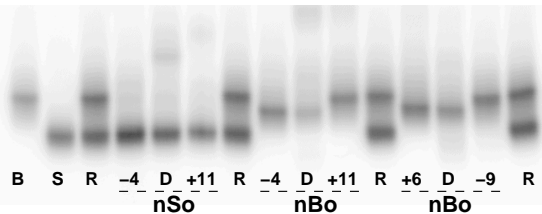,width=8.2cm,angle=0.}}
\caption{\label{F2n}
PAGE analysis of comparative mobilities of double nicked DNA
constructed from the random and A-tract repeat fragments. The contents
of the lanes is as follows (see \rtb{alseq} and Methods for fragment
codes). The lanes marked 'S,', 'B' and 'R' ("reference") contain the
intact fragments S, B, and their mixture, respectively. Lanes marked
by integer numbers contain the corresponding single-nicked DNA
fragments of the series marked below. Lanes marked 'D' contain DNA
fragments carrying simultaneously the two nicks from the left and right
neighbor lanes.}\end{figure}

The experimental results presented above evidence that single stranded
breaks really reduce the overall curvature of the A-tract repeat in
remarkable qualitative agreement with CBT. Other theories of DNA
bending cannot account for these data unless strong {\em ad hoc}
assumptions are introduced like sequence-and-position dependent nick
wedges, for example. 

\subsection{DNA Curvature in Simulations}

\begin{table}[t]\caption{\label{Tenpa} Structural parameters of
standard and computed DNA conformations. The top row features the
canonical B-DNA values corresponding to the fiber crystal
structure\cite{Arnott:72}. The RMSD values were computed with respect
to this conformation. The results are shown for the last 1 ns averaged
structures from the corresponding trajectories. The data for B and S
trajectories are taken from our previous report \cite{Mzpre:02}. The
helicoidals are the sequence averaged values computed with program
Curves \pcite{Curves:}. All distances are in angstr{\"o}ms and angles
in degrees.}
\begin{ruledtabular}
\begin{tabular}[t]{|cccccc|}
      & Xdisp
             & Inclin
                     & Rise
                           & Twist
                                  & RMSD \\
\hline
B-DNA & -0.7 & -6.0 & 3.4 & 36.0 &  0.0  \\

S     & -0.1 & -4.7 & 3.5 & 34.4 &  4.1  \\
B     & -0.4 & -4.0 & 3.5 & 34.2 &  6.8  \\
nB-4o & -0.4 & -3.3 & 3.5 & 34.7 &  6.0  \\
nB-2o & -0.3 & -3.4 & 3.5 & 34.5 &  3.2  \\
nB+0o & -0.1 & -5.0 & 3.5 & 34.5 &  6.4  \\
nB+2o & -0.7 & -4.4 & 3.5 & 34.1 &  4.0  \\
nB+4o & -0.6 & -4.5 & 3.5 & 34.1 &  3.3  \\
nB+6o & -0.6 & -4.3 & 3.5 & 34.3 &  4.8  \\
\end{tabular}
\end{ruledtabular}
\end{table}

\begin{table}[t]\caption{\label{Tnickpa} Local inter-base parameters
at strand breaks. The nicked steps only were
analyzed with Curves \pcite{Curves:} in the last five 1ns averaged
structures. The same steps were analyzed in their intact states by
similarly using the B trajectory of the mother A-tract fragment.
Each fragment is characterized by two lines showing the averages and
standard deviations for the given base step in its nicked (top) and
intact (bottom) states. All distances are in angstr{\"o}ms and angles
in degrees.}
\begin{ruledtabular}
\begin{tabular}[t]{|cccccc|}
Shift & Slide & Rise & Tilt & Roll & Twist \\
\hline
&&&$\underline{\rm nB-4o}$&&\\
 1.2$\pm$0.3&-0.4$\pm$0.1&3.3$\pm$0.1&  9.7$\pm$1.7&-14.6$\pm$ 4.9&40.5$\pm$1.3\\
 0.9$\pm$0.8&-0.7$\pm$0.2&3.2$\pm$0.2&  7.8$\pm$2.7& -4.1$\pm$ 5.1&37.5$\pm$5.8\\
                &&&$\underline{\rm nB-2o}$&&\\
 0.5$\pm$0.2& 2.6$\pm$0.1&3.2$\pm$0.1& 12.7$\pm$1.4&  4.2$\pm$ 3.0& 7.8$\pm$2.9\\
 1.2$\pm$0.9& 0.1$\pm$0.2&3.2$\pm$0.2& 10.3$\pm$3.4&  5.8$\pm$ 2.7&37.5$\pm$2.2\\
                &&&$\underline{\rm nB+0o}$&&\\
-1.0$\pm$0.1& 1.2$\pm$0.2&3.6$\pm$0.2&-11.1$\pm$6.5&  1.8$\pm$ 2.4&52.2$\pm$2.8\\
 0.4$\pm$0.2&-1.2$\pm$0.2&3.5$\pm$0.1&  0.7$\pm$1.0& -1.8$\pm$ 2.7&32.6$\pm$1.0\\
                &&&$\underline{\rm nB+2o}$&&\\
-0.9$\pm$0.1&-0.1$\pm$0.6&2.8$\pm$0.1& 14.0$\pm$2.7& -2.3$\pm$ 7.9&20.7$\pm$2.7\\
 0.1$\pm$0.2&-1.8$\pm$0.1&3.8$\pm$0.1& -2.6$\pm$2.6& -3.8$\pm$ 1.5&32.0$\pm$1.6\\
                &&&$\underline{\rm nB+4o}$&&\\
-0.6$\pm$0.2&-1.3$\pm$0.2&3.3$\pm$0.2&  5.3$\pm$2.8&-10.8$\pm$10.1&25.6$\pm$1.7\\
-0.2$\pm$0.1&-1.5$\pm$0.2&3.4$\pm$0.1&  3.6$\pm$2.1&-10.3$\pm$ 2.8&30.4$\pm$0.8\\
                &&&$\underline{\rm nB+6o}$&&\\
 0.8$\pm$0.1&-0.4$\pm$0.3&3.4$\pm$0.2&  9.6$\pm$2.0&-23.9$\pm$ 5.1&39.0$\pm$1.7\\
 0.5$\pm$0.2&-1.1$\pm$0.2&3.4$\pm$0.2&  3.7$\pm$3.0& -8.3$\pm$ 1.2&37.0$\pm$2.1\\
\end{tabular}
\end{ruledtabular}
\end{table}

As regards the overall stability of dynamics and its closeness to the
B form, all MD trajectories computed in the course of this studies
were similar to our earlier simulations under similar conditions
\cite{Mzjacs:98,Mzjcc:01,Mzjacs:00,Mzjbsd:01,Mzpre:02}. Table
\ref{Tenpa} shows parameters of the final 1ns-average conformations.
They all have remarkably similar average helicoidals corresponding to
a typical B-DNA. For example, the average helical twist estimated from
the best-fit B-DNA experimental values \cite{Kabsch:82} gives $34.0\pm
0.2$\degree\ and $33.8\pm 0.2$\degree\ for the A-tract fragment and
the randomized sequence, respectively. The nicks did not cause readily
visible perturbations and it was commonly difficult to distinguish
them in snapshots. Statistical analysis of the local inter-base
parameters at nick positions is summarized in \rtb{nickpa}.  As
expected, the broken DNA strand exhibits somewhat increased
flexibility at the nick site. The most noticeable are much stronger
deviations of Twist form its canonical values. A similar feature is
smaller, but distinguishable for Tilt and Roll. On the other hand, the
nanosecond time scale fluctuations characterized by the standard
deviations are not very different with and without the nick. The last
observation suggests that, as long as the bases at the nicked step
remain in the stack, their rapid motions are restrained to the same
degree as in the intact structure, and that the above strong
deviations of some averages are due to slower global motions that
occur in much longer time scales.  Apart from these strong
fluctuations, however, we did not observe any repetitive static
perturbations that might be attributed to strand breaking. Moreover,
strongly non-canonical helical parameters are sometimes encountered in
intact steps as well.  As it was for dynamics of the mother A-tract
DNA fragment\cite{Mzpre:02}, the rms deviations from the canonical
structures seen in \rtb{enpa} are largely due to variation of the
overall curvature.

\begin{figure}
\centerline{\psfig{figure=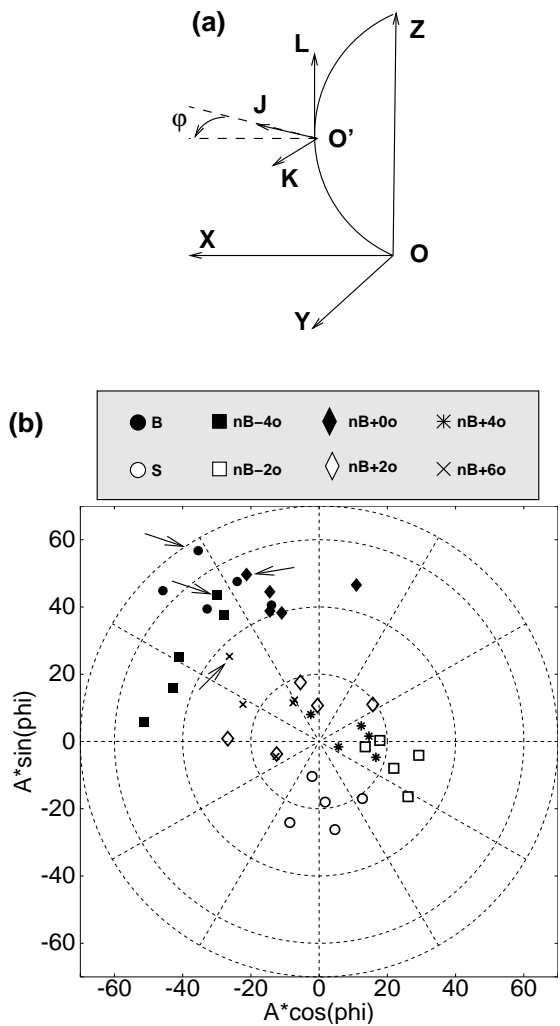,width=8cm,angle=0.}}
\caption{\label{Fpogr}
(a) Geometric constructions used for evaluating the DNA bending. The
amplitude of bending is measured by the angle between the ends of the
curved helical axis produced by the Curves algorithm \pcite{Curves:}.
The two coordinate frames shown are the global Cartesian coordinates
(OXYZ), and the local frame constructed in the middle point according
to the Cambridge convention (O'JKL) \pcite{Dickerson:89}. The curve is
rotated with two its ends fixed at the Z-axis to put the middle point
in plane XOZ. The bending direction is measured by angle $\varphi$
between this plane and vector $\bf J$ of the local frame. By
definition, this vector points to the major DNA groove along the short
axis of the reference base pair \pcite{Dickerson:89}. Consequently,
the zero $\varphi$ value corresponds to the overall bend towards the
minor groove in the middle of the DNA fragment as it was in the very
first analyses of local DNA curvature \pcite{Zhurkin:79}.\\ (b) Polar
plot of bending angle (A) versus direction (phi). The results are
shown for the last five 1 ns averaged structures from each trajectory,
with the curvature measured as explained in plate (a).}\end{figure}

\begin{figure}
\centerline{\psfig{figure=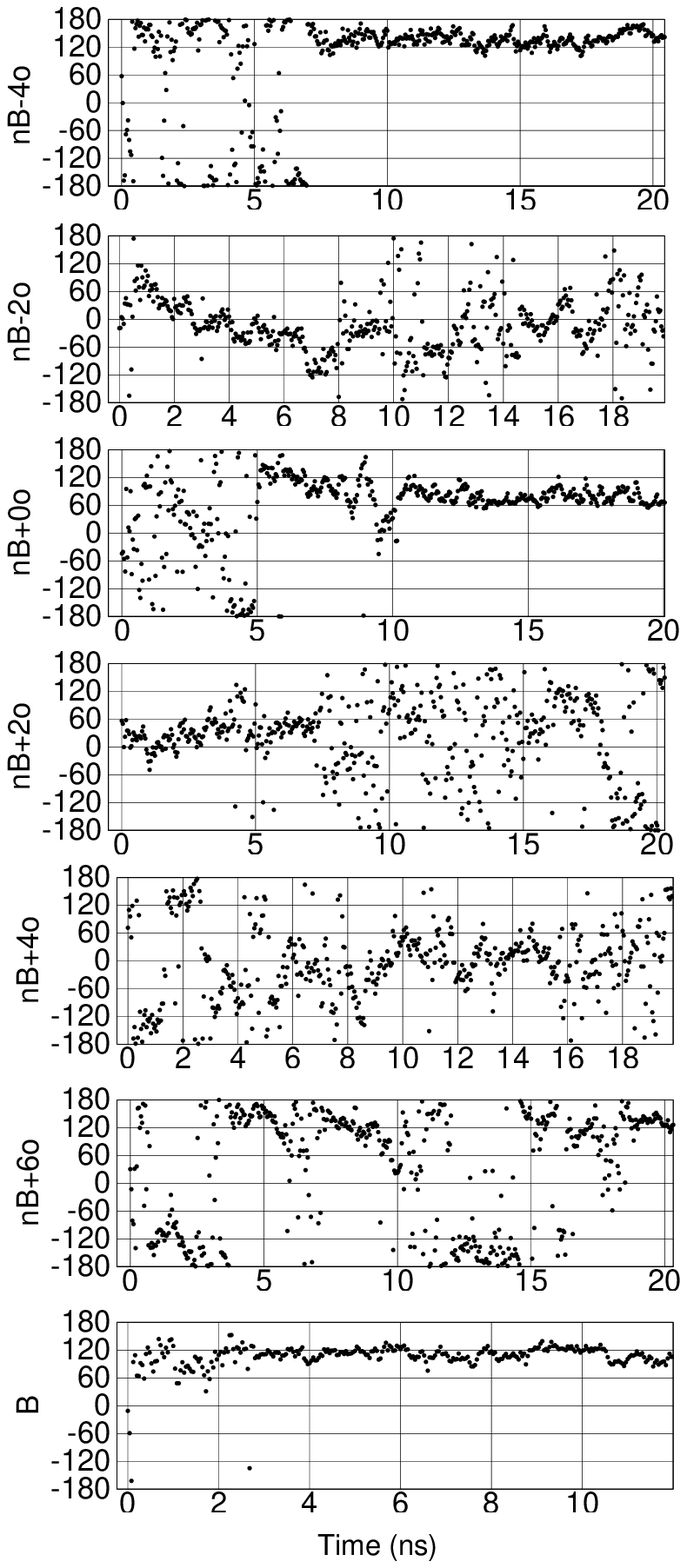,width=8cm,angle=0.}}
\caption{\label{Fbnddc}
The time evolution of the bending direction as measured by the
$\varphi$ angle in \rfg{pogr}a (in degrees). Here the axis of the
curved double helix was computed as the best fit common axis of
coaxial cylindrical surfaces passing through sugar atoms, which gives
solutions close to those produced by the Curves algorithm
\pcite{Curves:}. The traces have been smoothed by averaging with a
window of 60 ps and are displayed with a time step of 40
ps.}\end{figure}

\rfg{pogr} compares the overall DNA curvature in the last five 1ns
averaged conformations from all trajectories mentioned in \rtb{enpa}.
It is seen that the mother A-tract fragment (B) reached the largest
bending amplitudes non accessible to other fragments. The most stable
and significant curvature among the nicks is obtained for
nB-4o and nB+0o. Detailed in \rfg{bnddc} is the dynamics of the bend
direction. Its short time scale fluctuations are smaller when bending
is strong, therefore, \rfg{bnddc} also gives an estimate of the
bending amplitude. Again we note that the mother A-tract fragment (B)
is distinguished by the rapid establishment of curvature. However,
convergence to a statically bent state is clear for nB-4o and nB+0o as
well. Following to our previous
approach\cite{Mzjacs:00,Mzjbsd:01,Mzpre:02} an attempt has been made
to confirm the attracting property of the bent state for nB+0o by
running dynamics starting from the canonical A-DNA conformation. The
corresponding trajectory was continued to 17 ns and it showed
convergence of the same bending direction as in \rfg{bnddc}, but with
a somewhat smaller amplitude.

\rfg{pogr} and \rfg{bnddc} show that our MD simulations failed to
reproduce a regular variation of the bending magnitude with the nick
position as seen in the PAGE experiments above. Fragments nB-4o and
nB+0o exhibit noticeable curvature comparable, but smaller
than that of the intact duplex. According to experiment, the
curvature in nB+0o is similar to that in the mother fragment, but for
nB-4o it should be definitely smaller. Fragment nB+6o exhibited less
stable bending in a similar direction whereas for nB-2o, nB+2o, and
nB+4o the 20 ns duration of trajectories apparently was not sufficient
for convergence. At the same time, our calculations confirm that, in
qualitative agreement with experiment, single strand breaking
interferes with bending and generally reduces it. It is remarkable in
\rfg{pogr} that strong bends beyond 40\degree\ were observed only in
directions roughly corresponding to that in the mother fragment, which
confirms the strong attracting property of this bent state. Taking
into account the unclear physical nature of the phenomenon being
studied and the complexity of the model system we consider the partial
qualitative agreement achieved here as very satisfactory.

\begin{figure}
\centerline{\psfig{figure=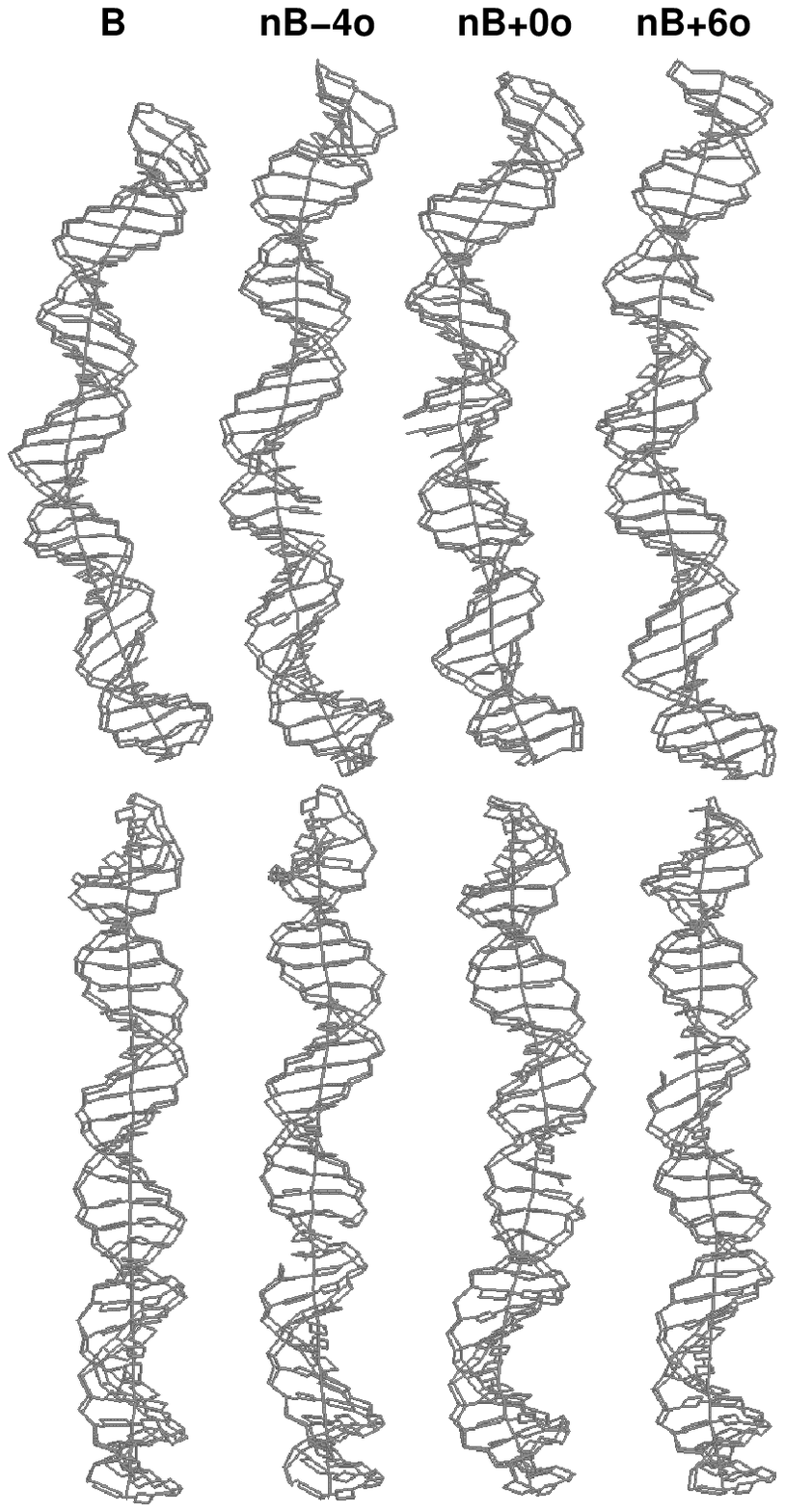,width=8cm,angle=0.}}
\caption{\label{F4bnd}
Schematic drawings of the of 1 ns averaged structures marked by arrows
in \rfg{pogr}. The pictures were produced with Curves\cite{Curves:}.
The backbone breaks are distinguished by three consecutive gaps in the
ribbons, but only the central phosphate is in fact absent. Two
orthogonal views are shown. The upper and lower views correspond to
the negative direction of the OY axis and the positive direction of
the OX axis, respectively, as shown in in \rfg{pogr}a.}\end{figure}

In spite of this only limited agreement with experiment, we tried to
use the ensemble of computed curved DNA conformations to clarify the
origin of the effect produced by nicks upon the A-tract curvature. To
this end, for each of the fragments B, nB-4o, nB+0o, and nB+6o a
nanosecond of dynamics have been selected for a more detailed
analysis. The corresponding 1ns average structures are marked by
arrows in \rfg{pogr}. They all correspond to the largest degree of
curvature reached during these trajectories. For B and nB+6o these are
the last nanosecond structures. In contrast, for nB-4o and nB+0o the
17th and the 19th nanoseconds were used, respectively.  According to
\rfg{pogr}, these bends are more or less convergent suggesting that
the corresponding dynamics should be similar and subtle differences
caused by nicks might be detectable. As seen in \rfg{4bnd}, for
fragments B, nB-4o, and nB+6o the averaged conformations are rather
close to one another. The curved DNA axis is almost planar with two or
three zones of increased bending found between the A-tracts. In
contrast, for nB+0o the curved axis is strongly non-planar due to
significant deviation of bend direction in its central zone from the
other two. As a result, for B, nB-4o, and nB+6o the overall bend is
close to the sum of local bends, but for nB+Oo it is much smaller. The
minor groove profiles averaged over the corresponding 1 ns periods are
compared in \rfg{mg}. Again one can note a similarity between B,
nB-4o, and nB+6o versus a more complex profile for nB+Oo.  The
widenings of the minor groove are found between A-tracts and they
roughly correspond to the zones of increased bending in \rfg{4bnd}.
According to \rfg{4bnd} and \rfg{mg}, a reduced overall bending
amplitude in nB-4o and nB+6o with respect to the mother A-tract
fragment is attributable to the central zone where the widening is
also less significant. In contrast, the smaller curvature in nB+0o is
mainly due to the lost phasing between the local bends.

Relaxation of A-tract curvature by single stranded breaks has been
originally proposed as a conceptual test of CBT\cite{Mzpre:02}. The
results reported here qualitatively agree with its predictions and
strongly support this model versus its alternatives. At the same time,
the nick position dependence as revealed by our PAGE tests is inverted
with respect to to the original guess\cite{Mzpre:02}. We thought that
the backbone compression is maximal in the widenings of the minor
groove, therefore, nicks between A-tracts should relax bending
stronger than nicks inside them. In contrast, the experiments reported
above suggest that the backbone compression should increase inside
A-tracts and reduce outside them. Because A-tracts are found at the
internal edge of the curved double helix the last suggestion agrees
with the simple physical intuition that says that the surface of a
curved cylinder is compressed at its inner edge and stretched at the
opposite side.  In order to understand which backbone component can be
responsible for the compression/frustration relationship postulated by
CBT we tried to measure the specific backbone length by using pairs of
similar atoms in neighboring residues.

\begin{figure}
\centerline{\psfig{figure=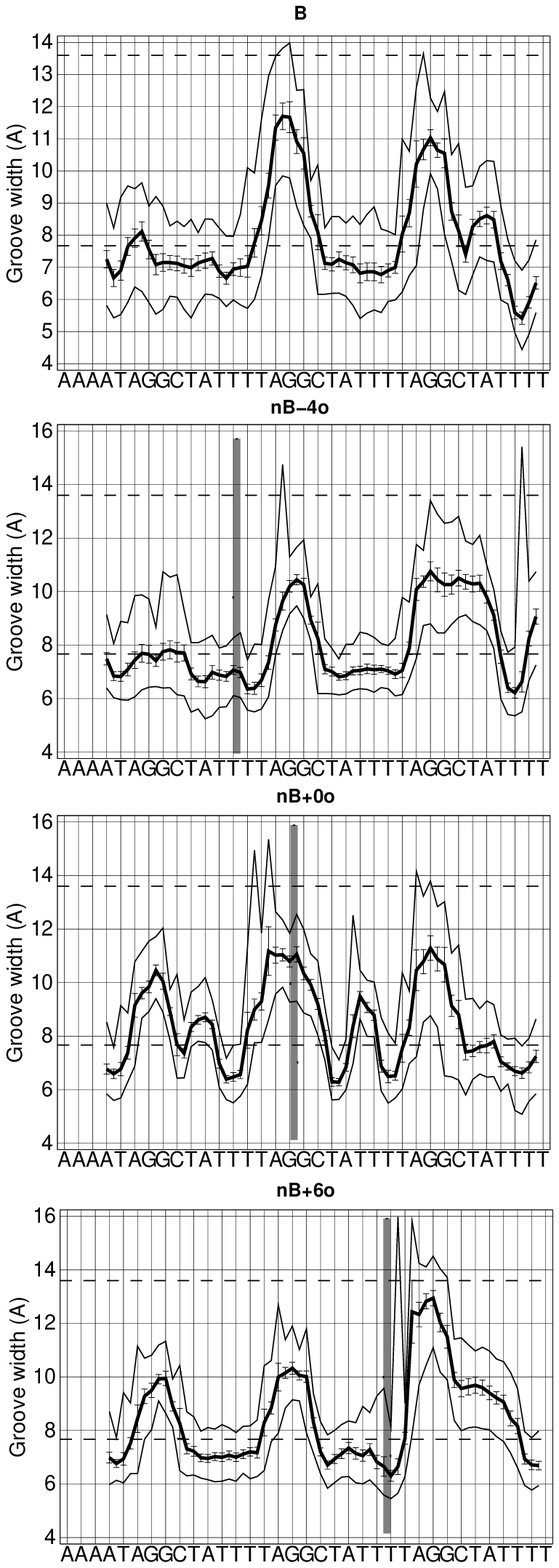,width=8cm,angle=0.}}
\caption{\label{Fmg} The profiles of the minor groove profile averaged
over the 1 ns trajectory intervals marked in \rfg{pogr}. The central
traces represent the average groove width with rms fluctuations shown
as error bars. The upper and lower solid traces show the maximal and
minimal values, respectively.  The groove width is evaluated by using
space traces of C5' atoms \pcite{Mzjmb:99}. Its value is given in
angstr\"oms, with the corresponding canonical A- and B-DNA levels
indicated by the horizontal dashed lines. The vertical grey bars mark
nick positions.}\end{figure}

\begin{figure}
\centerline{\psfig{figure=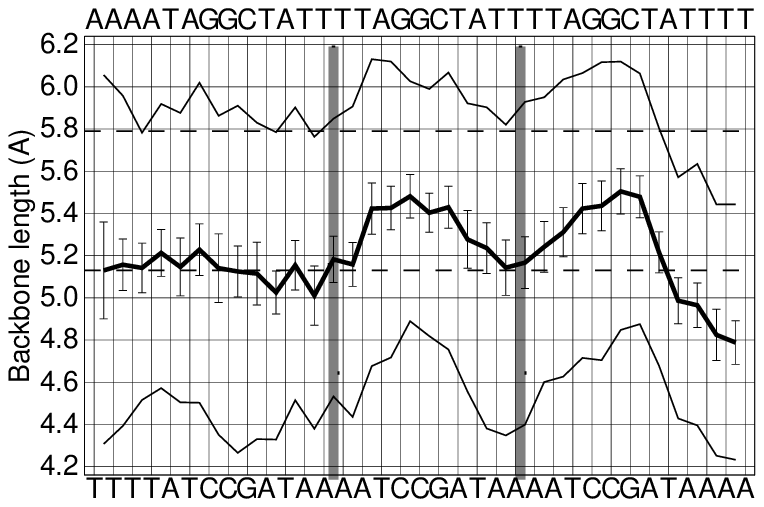,width=8cm,angle=0.}}
\caption{\label{Fsgdi}Backbone length profiles averaged over the 1 ns
trajectory intervals marked in \rfg{pogr}.  The plots were generated
by using inter-O4' distances.  The central traces represent the
average with rms fluctuations shown as error bars. The upper and lower
solid traces show the maximal and minimal values, respectively. All
profiles were smoothed with a sliding window of 3 base pair steps. The
length is given in angstr\"oms, with the corresponding canonical A- and
B-DNA levels indicated by the horizontal dashed lines. The vertical
grey bars mark nick positions.}\end{figure}

For all atom pairs the resulting profiles of the backbone length appear
very noisy and regular modulations can be seen only after averaging
with a sliding window. Inspection of all atom pairs allows one to
divide them in two groups. For atoms of phosphate groups and their
neighbors, the measured backbone length usually is larger at the inner
side of the bend and smaller at the opposite side, which may be viewed
as a trend towards the A-from in the widenings of the major groove.
However, these modulations are very weak, in agreement with the
conventional view of the B-DNA backbone as being non-compressible
\cite{Packer:98,Calladine:92}. In contrast, for a group of backbone
atoms including O4' and its close neighbors, strong regular
oscillations of the backbone length could be revealed, with their
phases being opposite to those in the first group.  The average
backbone profiles produced by inter-O4' distances are shown in
\rfg{sgdi}. They were computed for the 1 ns intervals of trajectories
marked in \rfg{pogr}. It is seen that the backbone length correlates
with the minor groove profiles in \rfg{mg}, but does not repeat them.
Thus measured specific backbone length reaches its local minima at the
inner edge of the bent DNA cylinder and corresponds exactly to the
nick positions in nB-4o and nB+6o.  Similar oscillations are observed
in X-ray structures of bent DNA available in NDB\cite{NDB:} (AKM,
unpublished). It may seem surprising that profiles in \rfg{sgdi}
exhibit no perturbations at strand breaks, but this results from
smoothing. In the non-smoothed plots the single stranded breaks in
nB+6o and nB+0o were distinguished by very small and very large
distances, respectively, but they have been compensated by strong
opposite deviations in the neighboring steps.

The foregoing analysis demonstrates that the DNA backbone behaves as a
complex mechanical system rather than a chain of strings or rigid
rods, and that different its components compress and stretch in
different zones of the curved DNA. Nevertheless, the correlations
revealed in \rfg{sgdi} suggest that the backbone compression, if it
really exists, should mainly affect sugar-sugar interactions and that
they can be at the origin of the intrinsic backbone frustration.  Even
though in B-DNA consecutive O4' atoms do not interact, the neighboring
sugar rings make a number of direct contacts non-mediated by the
solvent.

\section{Concluding Discussion}

The results presented here provide new insights in the putative
physical mechanism of intrinsic DNA curvature. This intriguing
phenomenon seems to be exhaustively studied, but it still attracts
great interest. The A-tract repeats and single stranded breaks both
have been earlier recognized as elements specifically involved in DNA
bending. To our best knowledge, however, nobody checked what could be
their cumulative effect. Here it is found that nicks relax the
intrinsic curvature induced by A-tract repeats in conditions where
they do not affect the random sequence DNA. The relaxation effect
depends upon the nick position with respect to the bend. It is very
significant when the nick is at the inner edge of the bent DNA inside
A-tracts, and is gradually reduced essentially to zero when the nick
is moved to the outer surface of the bend. This behavior is in
qualitative agreement with the CBT and it suggests that the backbone
compression in the curved DNA is larger in the minor groove narrowings
at the inner surface.

The MD simulations confirm that single stranded breaks interfere with,
and tend to relax the A-tract induced curvature.  For about a half of
the DNA fragments studied, the duration of trajectories provided
reproducible convergence to statically bent states in good agreement
with experiment as well as earlier simulations. Inspection of the
computed conformations reveals strong regular modulations of the local
backbone length as measured by distances between certain sugar atoms,
with compression reaching its maximum at the inner edge of the bend
where nicks produce the strongest relaxation in experiments. These
results provide additional support to the CBT and suggest that the
frustration in the B-DNA backbone may result from interactions between
consecutive sugar rings.

\clearpage
\bibliography{preprint}
\bibliographystyle{apsrev}

\end{document}